\documentclass[twocolumn]{aastex62}

\usepackage{amssymb}
\usepackage{amsmath}
\usepackage[]{graphicx}
\usepackage{enumerate}
\usepackage{mathrsfs,amssymb}

\usepackage{color} 
\usepackage{xspace}

\tightenlines

\begin{document}

\title{
Analytical and numerical studies of central galactic outflows powered by tidal disruption events \\
- a model for the Fermi bubbles?
}

\author{C. M. Ko}\altaffiliation{cmko@astro.ncu.edu.tw}
\affiliation{Institute of Astronomy, National Central University, Zhongli Dist., Taoyuan City, Taiwan (R.O.C.)}
\affiliation{Department of Physics and Center for Complex Systems, National
Central University, Zhongli Dist., Taoyuan City, Taiwan (R.O.C.)}
\author{D. Breitschwerdt}
\affiliation{Zentrum f\"ur Astronomie und Astrophysik, Technische Universit\"at Berlin, Hardenbergstrasse 36, 10623 Berlin, Germany}

\author{D. O. Chernyshov}
\affiliation{I.E.Tamm Theoretical Physics Division of P.N.Lebedev Institute of Physics, Leninskii pr. 53, 119991 Moscow,
Russia}
\affiliation{Moscow Institute of Physics and Technology (State University), 9, Institutsky lane, Dolgoprudny, 141707, Russia}
\author{H. Cheng}
\affiliation{Institute of Astronomy, National Central University, Zhongli Dist., Taoyuan City, Taiwan (R.O.C.)}

\author{L. Dai}
\affiliation{Department of Physics, The University of Hong Kong, Pokfulam Road, Hong Kong}
\affiliation{DARK Cosmology Centre, Niels Bohr Institute, University of Copenhagen, Juliane Maries Vej 30, 2100 Copenhagen O, Denmark}

\author{V. A. Dogiel}\altaffiliation{dogiel@td.lpi.ru}
\affiliation{I.E.Tamm Theoretical Physics Division of P.N.Lebedev Institute of Physics, Leninskii pr. 53, 119991 Moscow,
Russia}




\begin{abstract}
Capture and tidal disruption of stars by the supermassive black hole in the Galactic center (GC) should occur regularly.
The energy released and dissipated by this processes will affect both the ambient environment of the GC and the Galactic halo.
A single star of super-Eddington eruption generates a subsonic outflow  with an energy release of more than $10^{52}$ erg,
which still is not high enough to push shock heated gas into the halo.
Only routine tidal disruption of stars near the GC can provide enough cumulative energy to form and maintain large scale structures like the Fermi Bubbles.
The average  rate of disruption events is expected to be $10^{-4}\sim 10^{-5}$ yr$^{-1}$, providing the average power
of energy release from the GC into the halo of $\dot{W}\sim 3\times 10^{41}$ erg s$^{-1}$, which is needed to support the Fermi Bubbles.
The GC black hole is surrounded by molecular clouds in the disk, but their overall mass and filling factor is
too low to stall the shocks from tidal disruption events significantly.
The de facto continuous energy injection on timescales of Myr will lead to the propagation of strong shocks in a density stratified Galactic halo
and thus create elongated bubble-like features, which are symmetric to the Galactic midplane.
\end{abstract}


\section{Introduction}
\label{sec:introduction}

Two enigmatic gamma-ray features in the Galactic central region
also known as Fermi Bubbles (FBs) were found from Fermi-LAT data \citep[see][]{dob10,su10,acker14}.
The X-ray and microwave emission around the Galactic center (GC) was detected by \citet{Snowden1997,bland03} and
\citet{fink04}.
Later observations from Planck \citep[see][and publications in \citealt{rub18,jew20}]{planck13}
showed structures coincided nicely with the FBs.

These features elongated perpendicular to the Galactic plane, is seen as a double-bubble structure on the two sides of the plane.
The spatial distribution of the emissions from the bubbles shows sharp edges.
The surface emissivity is almost uniform inside the bubbles.

These characteristics of gamma-ray and microwave emissions may naturally be interpreted by the radiation of relativistic electrons
which are accelerated in-situ near the bubble edges \citep[see, e.g.,][]{su10,mertsch2011,cheng11,cheng14,cheng15}.
\citet{Crocker2011} suggested an alternative model of gamma-ray emission from the bubbles which is produced by collisions of relativistic protons.
The reader is referred to \citet{yang2018} for more details.

We do not consider the problem of particle acceleration in the FBs but concentrate on the origin of their structure and their hydrodynamic evolution.
Although the nonthermal and thermal envelopes are of the same origin in general,
their structures are defined by different physical processes and their interconnection is indirect.

The ROSAT 1.5 keV image presented a giant structure in the Galactic center
which was seen in thermal X-rays as an open-ended bipolar hour glass \citep[see][]{bland03}.
This phenomenon required a central event with energetics of about $10^{55}$ ergs whose activity had a timescale of $10\sim 15$ Myr.

Central galactic outflows have been observed in several galaxies.
For example, a structure similar to the FBs was reported recently in NGC\,3079
in both non-thermal hard X-rays (Chandra) and in non-thermal radio continuum (JVLA) \citep{li19}.
The hard X-ray feature has a cone shape with a weak cap at the top with a diameter $\sim 1.1$ kpc.
The authors have found evidence for a significant thermal component above 1 keV,
and also argue that cosmic ray electrons have to be accelerated in situ by shocks.
Since most of the data, from gamma to X-rays to radio have been collected for our own galaxy,
we will focus here on the so-called Fermi Bubbles.

%
%

The origin of energy release at the GC is still an open question.
One possibility is that
the initial energy release at the GC is a result of accretion onto the central supermassive black hole (SMBH) with a mass of $\sim 4.3 \times10^6~M_{\odot}$,
which is identified with the source Sgr A* \citep[see][]{gil09}.
The total energy needed to generate large Galactic outflows is assumed to be in the range
from $10^{53}$ to $10^{56}$ erg  \citep[see][]{bland03,su10,akita18,kesh18}
if the initial energy release occurred close to the GC in the past.

Several models of the FBs involve a gas outflow from Sgr A* after a  huge energy release at the GC ($\sim 10^{55} - 10^{56}$ erg).
For example, \citet{zub11,zub12} \citep[see also][]{nayak18} suggested that
a giant molecular cloud of mass $\sim 10^5M_\odot$ was captured by the SMBH in the GC about one Myr ago.
A simplified model is described by a shock front of a relatively short single burst propagating through  the halo of a uniform density.

\citet{yuan14} assumed an alternative scenario, in which the mass accretion rate of hot gas flow onto Sgr A* was $10^3 \sim 10^4$ times higher in the past.
Its activity lasted for $10^7$ yr and ceased about $2\times 10^5$ yr ago.
During these $10^7$ years the bubble propagated several kiloparsecs out into the halo, driven by winds launched from the accretion flow onto Sgr A*.
Similar models of ongoing energy release in the GC were suggested for the interpretation of the local GC $e^\pm$ annihilation line
\citet{cheng06,cheng07} and \citet{tota06}.

\citet{guo12,guo12a,zhguo20} \citep[see also][]{yang2012} developed a model of the bubble at the GC with a recent active galactic nucleus (AGN) jet activity,
occurring $1\sim 3$ Myr ago.
Cosmic ray effects were included in their MHD model.
The FB evolution is described by a system of nonlinear hydrodynamic equation, similar to \citet{drury} for CR acceleration at shocks
or to \citet{brei91} for wind escape from the Galactic wind.
The active period of energy release at the GC is $0.1\sim 0.5$ Myr. The total energy release was estimated as $10^{55} \sim 10^{57}$ erg.
The model of the envelope with arbitraty/undefined parameters describes, nevertheless, a proper structure of the FBs.

We notice that sharp edges of the FBs and the uniform surface emissivity inside the bubble can be interpreted as a result of acceleration of relativistic electrons
by turbulence and shocks inside and near the boundary of FBs \citep[see][]{cheng11,cheng14,cheng15}.

An alternative energy release was suggested in the GC by star formation activity over about $10^7$ years \citep[see e.g.][]{carr13,naka18,zhang20}.
In principle this process may also produce an
energy outflow by star formation region in the CMZ
(central molecular zone, a sheet like structure surrounding the GC).
We note that star formation activity near the GC is still under debate.
Some studies of star formation regions near the GC indicated that star formation activity is suppressed,
and it may not play a significant role in comparison with processes in starburst galaxies \citep[see, e.g.,][]{kauf17}.
On the other hand, there are indications that there might be increased star formation activities at GC in recent time \citep[][]{genz10,Nogueras-Lara2020}
(although not comparable to starburst galaxies).

Recent X-ray observations in the direction of the constellation Draco (X-ray source Swift J1644+57) found a more moderate energy release.
The peak luminosity was detected to be  $\sim 10^{48}$ erg s$^{-1}$, and the total energy release was estimated to be $3\times 10^{53}$ erg \citep{bur11}.
Another example of a huge energy release was presented in \citet{donato14},
who interpreted a flare in the cluster Abell~1795 as a stellar disruption with an energy release of about
$1.7\times 10^{52}$ erg by a black hole with a mass  of $\sim 3\times 10^5M_\odot$.
\citet{li20} presented results of monitoring observations of a stellar tidal disruption event (TDE)
by a supermassive black hole ($\sim 5 \times10^7$ M$_\odot$) in NGC5092.
Over a period of 13 years the overall X-ray luminosity was estimated to be $\sim 1.5 \times 10^{43}$ erg s$^{-1}$.

\citet{cheng11,cheng12} speculated about the origin of the Fermi Bubbles
as the result of routine stellar tidal disruption events by the central black hole.
The expected energy release produced by accretion processes in the GC is about $\ga 10^{52}$ erg,
depending on the mass of the captured stars close to Sgr A. Less than 0.04 pc from it are about 35 low-mass stars ($1\sim 3 M_\odot$)
and about 10 massive stars ($3\sim 15 M_\odot$), see, e.g.  \citet{alex04, alex05, genz10}.
The expected rate of stellar capture per galaxy from theoretical calculation is $\sim 2\times0^{-4}$ yr$^{-1}$  \citep{mag99,wang04,stone2018}.

Observational evidence of energy release from Sgr A* was found recently.
The estimates are very crude, but they give a qualitative understanding of processes involved.
X-ray and radio observations showed a pair of lobes on a scale of about 15 pc, located above and below the Galactic plane,
surrounding the GC \citep{morris2003,zhao2016}.
They are filled with plasma at temperature $0.7\sim 1$ keV.
More recently, using XMM and Chandra, \citet{ponti19} obtained a detailed X-ray map of a region nearby Sgr A*.
They found two elongated structures extending above and below the GC,
which are called the northern and southern Galactic Centre chimneys.
These quasi-linear structures are about 160 pc in length and have sharp edges.
Their thermal energy content is about $4\times 10^{52}$ erg.
The gas density within the chimneys decreases with latitude from about 0.2 cm$^{-3}$ at 30 pc to about 0.1 cm$^{-3}$ at 160 pc.
The authors suggested that the chimneys connect the regions around the GC to the FBs.
An X-ray plume of size $\sim 1^\circ$ observed by Suzaku is interpreted as a magnetized hot gas outflow from the GC
\citep{Nakashima2019}.

Using the MeerKAT, \citet{hey19} found a pair of radio bubbles at the GC.
The structure is seen as a pair of bounded bipolar bubbles spanning $140\,{\rm pc} \times 430\,{\rm pc}$ across the Galactic plane.
 The radio emission is consistent with synchrotron radiation.
The total energy in the radio bubbles is estimated to be $7\times 10^{52}$ erg.
The energy is much less than the total energy content of the FBs, but the authors indicated that
the radio bubbles may be an example of a series of similar events (and possibly combined with steadier outflows),
where cumulative effects may be responsible for the radio, X-ray and gamma-ray structures that connect the GC to the halo.

These observations provide evidence for energetic outbursts from the surroundings of the GC,
which propagate preferentially perpendicular to the Galactic plane, and, in combined action, may produce the necessary amount of energy to generate the FBs
or similar structures.

The paper is organized as follows.
In Section~\ref{sec:plasma} we discuss the energetics of TDEs at the GC.
In Section~\ref{sec:intohalo} we argue how the energy from the TDEs propagate through the Galactic disk to the halo.
Two models for the density distribution of the halo are described.
In Section~\ref{sec:analytical} we present analytical solutions for the shock propagation in the halo of the two halo models.
In Section~\ref{sec:simulation} we perform numerical calculations for the development of the bubbles in the two halo models.
Section~\ref{sec:summary} provides a summary.

\section{Plasma outflow generated by stellar disruptions and subsequent shock formation}
\label{sec:plasma}

In the following we adopt the hypothesis that cumulative routine stellar disruptions by the SMBH are responsible for the formation of the FBs
or similar structures.

X-ray observations of jetted tidal disruption events (TDEs) Sw 1644 \citep[][]{kara16} and non-jetted TDEs \citep{lin17}
can be interpreted as fast outflows from super-Eddington accretio. This process was analysed in numerical simulations by
\citet{dai18}.
They showed that the accretion energy is mainly carried away by the following channels:
(1) radiation with efficiency of $\eta_{\rm rad} \approx 3\%$,
(2) jet with $\eta_{\rm jet} \approx 20\%$, and
(3) outflow with $\eta_{\rm of} \approx 20\%$ (where $\eta$ is the ratio of the energy conversion rate to the rest mass accretion rate $\dot{M} c^2$).
Also, the outflow has a speed of several tenths of the speed of light, $c$, for most inclination angles.
The specific values of the efficiencies and the outflow speed depend on parameters such as the mass, the spin of the black hole and the accretion rate.
However, most of the current simulations of super-Eddington accretion with parameter settings have given consistent results
\citep[e.g.,][]{jiang,sadowski}.
Therefore, we adopt that in all tidal disruption events (TDEs),
outflows with speeds of several tenths of $c$ are produced, and that they carry away up to 10\% of the accretion energy.
Such fast outflows provide the largest impact on the surrounding matter since they can carry a lot of matter moving with non-relativistic speed.
Also, it is likely that only a small fraction of TDEs can produce jets.

Given the average mass of outbound matter to be approximately $0.5M_\odot$ for a TDE of a solar mass star, one can estimate the total energy of
the outflow as $\sim 10^{52} - 10^{53}$ ergs.
With these input parameters (outflow velocity $v_0 \approx 0.1c - 0.3c$, the total energy $W \approx 10^{52} - 10^{53}$ erg),
we intend to describe the outflow into the surrounding medium of the black hole.
In terms of total efficiency, it accounts for up to $1 \sim 10\%$ of the total rest mass energy of the disrupted star.

\section{Shock Propagation through the Galaxy}
\label{sec:intohalo}

\subsection{Shock Propagation through the Galactic Disk}
\label{sec:throughdisk}

A sudden (sporadic) energy release by a TDE in the GC  generates a cavity with a shock which expands into the local ambient medium.
This process is analogous to a supernova (SN) explosion and the subsequent evolution of the  supernova remnant (SNR).
The model of SNR shock propagation in a low pressure environment was developed by e.g. \citet{cox72,blond98}.
At the initial phase the expanding cavity can be described by the Sedov solution \citep{sed59}
when the total energy released is almost confined within the envelope.
In a later phase the solution differs from the similarity one due to radiative cooling in the envelope, and energy losses become important.
The pressure behind the neutral shell drops significantly, and the envelope expands until the pressure of the cavity reduces to the ambient
pressure of the interstellar medium (ISM), and the shell merges with the ISM.

The radius and time of transition from the Sedov phase to the radiative phase can be estimated as
\begin{equation}\label{eq:Rs}
R_s\approx 19.1\times W_{51}^{5/17}n_H^{-7/17}~\mbox{pc} \,,
\end{equation}
\begin{equation}\label{eq:ts}
t_s\approx 2.9\times 10^4\times W_{51}^{4/17}\, n_H^{-9/17}~\mbox{yr} \,,
\end{equation}
where $W_{51}$ is the energy release of TDE in units of $10^{51}$ erg and $n_H$ in cm$^{-3}$.

The question is whether the envelope pierces through the Galactic disk without significant braking,
so that the shock can pick up speed in a decreasing halo density environment.
The interstellar medium in the CMZ
is very complex.
The gas density in the ambient medium of the GC is not well known and quite difficult to assess.
A standard model of the gas distribution in the CMZ was presented by \citet{ferriere01,ferriere07,ferriere12a},
where about 80\% of the volume  consists of two phases: a hot coronal component at a temperature of about $10^5 \sim 10^6$ K
with an average density of about $\sim 10^{-3}$ cm$^{-3}$ and a warm ionised medium with a temperature of $T\sim 10^4$ K
and a density of $0.1 \sim 1$ cm$^{-3}$.
For a single explosion (i.e., a burst of energy)
in the GC with an energy release of $E_0\sim 10^{52} - {10}^{53}$ ergs,
the envelope still retains a fair amount of the blast wave energy in the Sedov phase for a single event,
when the Sedov radius $R_{ad}$ in the GC is about 100 pc, i.e. comparable or larger than the thick Galactic gas disk.
The cavity can then penetrate into the halo within its Sedov phase, even for a single explosion of moderate energy release,
and the radiation losses can be neglected.
One should also keep in mind that once gas has been pushed aside by an explosion,
a hole in the gas layer considerably alleviates the break-out of successive explosions.
The ``self-healing'' time of these holes is of the order of the cooling time and the subsequent loss of pressure,
plus the sound crossing time in which denser and cooler material flows in, which is of the order of $10^6 \sim 10^7$ yr,
much longer than the time interval for the next event to occur.

Recent observations of molecular lines in the directions of molecular clouds found an intermittent gas with average density about
$10^4$ cm$^{-3}$ \citep{mills17,mills18}.
However, the volume filling factor of molecular clouds in the CMZ is much less than 0.1.

\citet{oka19} suggested a new interpretation of the CMZ, unlike the standard model of the interstellar medium in the GC.
The volume of the CMZ is dominated by the diffuse molecular gas with a temperature $T\sim 200$ K and a density of $n\sim 50$ cm$^{-3}$.
The filling factor for dense gas with $n > 10^4$ cm$^{-3}$ should therefore be much less than 0.1.
The ultra hot X-ray-emitting plasma with a temperature $T\sim 10^8$ K, which some thought to dominate the CMZ,
does not coexist with the diffuse molecular gas and is thus not spread over extended regions.
The observed diffuse X-ray emission in the CMZ must be due to unresolved point sources and to scattering by interstellar atoms and molecules.
If this is true, the Sedov radius in the CMZ is about 15 pc for a single explosion of $10^{53}$ erg.
The cavity is unable to penetrate through the Galactic disk ($\sim 100$ pc), and its energy is dumped there.
The time of transition from the Sedov phase to the radiative phase for $n_H=50$ cm$^{-3}$ is about $t_s \simeq 1 \sim 2\times 10^4$ yr (see Eq.~(\ref{eq:ts})).
The  energy is transformed into radiation completely for  the time $t_E \sim 10^7$ yr within the radiative radius $R_E \sim 400$ pc \citep[see,][]{mckee},
i.e., a significant fraction of the energy release is lost in the disk.

Alternatively, if another TDE occurs within the time $t<t_s$ in the GC ,
then the cumulative effect provides a more extended Sedov radius than that of a single event.
The rate of the average tidal disruption events of stars is about $10^{-4}$ yr$^{-1}$ \citep{syer99},
and  the total average power in the GC can be estimated as $\dot{W} \la 3\times 10^{41}$ erg s$^{-1}$.
This scenario holds, when routinely occurring TDEs will punch through the disk and deposit energy into the halo.
However, we point out that the epoch of routine TDEs at the GC cannot exist forever.
It depends on the activities at the Galactic central regions,
which are highly variable
\citep[see, e.g.,][]{melia2001,genz10,Nogueras-Lara2020}.

This
type of
energy release is similar to processes of stellar winds described in \citet{aved72,cast75,weaver77}.
The hydrodynamic structure of the expanding cavity can be described by a Sedov solution for multi-captures.
The end of the adiabatic phase occurs at radius and time \citep{aved72},
\begin{equation}\label{eq:Rad}
R_{\rm ad}\approx 1.93\times \dot{W}_{36}^{2/5}n_H^{-3/5}~\mbox{pc} \,,
\end{equation}
\begin{equation}\label{eq:tad}
t_{\rm ad}\approx 7.5\times 10^3\times \dot{W}_{36}^{1/3}n_H^{2/3}~\mbox{yr} \,,
\end{equation}
where $\dot{W}$ is the power input in $10^{36}$ erg s$^{-1}$.

Even for $n_H\sim 50$ cm$^{-3}$ and a power in the GC of about $\dot{W} \la 3\times 10^{41}$ erg s$^{-1}$ the Sedov radius in the disk is 70 pc.
The thickness of CMZ is a few tens of parsecs \citep{morris96},
and for these parameters of the disk the envelope can penetrate freely into the halo, even for the case of multi-captures.

\subsection{Gas Distribution in the Halo}
\label{sec:halogas}

As in the disk, the gas distribution in the halo is not very reliable.
For instance, \citet{cordes,biswas} assume that the plasma density in the Galactic halo  drops exponentially with height $z$ above the Galactic plane,
\begin{equation}\label{eq:exp-model}
  n(z)=n_0\exp\left(-\frac{z}{z_0}\right)\,,
\end{equation}
where $n_0$ is the gas number density at $z=0$ and $z_0$ is the density scale height.
\citet{nord92} estimated the density of free electrons above the plane as $n_0=0.033$ cm$^{-3}$
and the characteristic scale of the electron distribution there as $H= 0.53 - 0.84$ kpc.
Similarly, \citet{gaen08} derived the warm plasma distribution ($\sim 10^4$ K) above the Galactic disk
($\la 2$ kpc with the average density $\sim 0.014$ cm$^{-3}$)
from pulsar dispersion measures and H$\alpha$ diffuse emission.

The gas distribution of the hot gas in the halo with temperature $T\sim 10^6$ K was derived from
Suzaku observation \citep{naka18}.
The distribution can be expressed as a disk-like density distribution
\begin{equation}
n(R,z)=n_0\exp\left(-\frac{R}{R_0}\right)\exp\left(-\frac{z}{z_0}\right)\,,
\end{equation}
with $n_0\simeq 4\times 10^{-3}$ cm$^{-3}$, the scale height $z_0\simeq 2$ kpc, and the radial scale length $R_0\simeq 7$ kpc.

The first indication on the X-ray structure was found by \citet{Snowden1997} from ROSAT, which was seen as a bulge of hot gas
similar to a cylinder with an exponential fall-off of density with height above the plane.
The cylinder has a radial extent around 5.6 kpc, and the scale height of 1.9 kpc with electron density at the base about
$0.0035$ cm$^{-3}$ and temperature about $10^6$ K.
Recent observations by Swift and Suzaku were interpreted as a bubble-in-halo in which two identical bubbles expand within a halo
forming a thick uniform shell of swept-up halo gas.
Assuming that the 0.3 keV plasma is heated by a shock driven by the bubbles' expansion in the surrounding halo, the corresponding velocity
is about $300$ km s$^{-1}$ \citep{Tahara2015,Kataoka2015}.

\citet{mill13,mill16} derived a so-called  $\beta$-model of the gas density profile in the halo from the intensity of absorption lines.
The diffuse gas density in the Milky Way halo is approximated by a flattened profile
\begin{equation}\label{eq:beta-model}
  n(R,z)=n_0\left[1+\left(\frac{R}{R_c}\right)^2 + \left(\frac{z}{z_c}\right)^2\right]^{-3\beta/2}\,,
\end{equation}
where $n_0\simeq 0.5$ cm$^{-3}$ is the core density, $R_c$ and $z_c$ represent the effective core radial and vertical distance,
and the exponent $\beta$ is the slope of the profile. For the GC gas distribution, we can use $R=0$, so that for $z\gg z_c$ the distribution simplifies to
\begin{equation}
  n(z)=n_0\left(\frac{z_c}{z}\right)^{3\beta} \,,
\label{beta_atm}
\end{equation}
and $z_c = 0.26$ kpc is used \citep[see][]{mill13}.

In these publications the authors presented a number of arguments in favour of the $\beta$-model \citep[see for details][]{mill13,mill16}.


It is clear that the exponential model, in comparison to the $\beta$-model, would overestimate the ability of shocks to penetrate the Galaxy halo,
as well as the distance of shock propagation into the Galactic halo, thus affecting the predicted bubble morphology as well.


\section{Analytical Solution for Shock Propagation in Exponential and Power-law Density Profile of the Galactic Halo}
\label{sec:analytical}

\citet{komp60} (see also \citet{shap79}, the review of \citet{kogan} and the monograph of \citet{zeld67})
developed the formalism of strong explosions with energy $E$ when a shock front is propagating through
an exponential atmosphere with the density distribution $n(z)=n_0\exp(-z/z_0)$.
He showed in this model that the shock propagates to infinity within a finite time.

\citet{baumbr13} developed an analytical solution of a hydrodynamic model for a shock wave propagation
for different energy input rates for single and successive explosions in exponential atmospheres in star forming regions in the disk.
The authors showed that the shock surface accelerates when its velocity is above the sound speed,
and that it develops Rayleigh-Taylor instabilities at the top of the envelope.
We expect that the shock propagation in a galactic halo with a $\beta$-atmosphere differs noticeably
from that of an exponential model in star formation regions or TDEs at GC.

Following the Kompaneets formalism \citep[see details in][]{kogan}, the shock front is described as
\begin{equation}\label{eq:kompaneets}
  \left(\frac{\partial r}{\partial y}\right)^2-\frac{1}{{\cal R}(z)}\left[\left(\frac{\partial r}{\partial z}\right)^2+1\right]=0\,,
\end{equation}
with a transformed time variable
\begin{equation}\label{eq:timevarible}
  y=\int\limits_0^t \sqrt{\frac{(\gamma^2-1)}{2}\frac{2W(t)}{3\rho_0V(t)}}\,dt\,,
\end{equation}
where $\gamma$ is ratio of the specific heats, $\rho_0=n_0 \bar{m}$ is the mass density of the background gas,
with $\bar{m}$ being the average atomic mass for a given metallicity, $W(t)$ is the energy released by a central source,
and $V(t)$ is the current volume of the expanding bubble,
\begin{equation}\label{eq:volume}
  V(t)=\pi\int\limits_{0}^{z_u}r^2(z,t)\,dz\,.
\end{equation}
Here $r(z,t)$ is the radius of the envelope in cylindrical coordinates, and $z_u$ is height of the top of the bubble.
${\cal R}(z)$ describes the structure of the atmosphere or halo, e.g., $\rho(z)=\rho_0 {\cal R}(z)$.
In the following we discuss two different structures:
(1) ${\cal R}(z)=\exp\left(-z/z_0\right)$, and (2) ${\cal R}(z)=\left(1+z/z_c\right)^{-2}$.

\subsection{Exponential Density Profile}
\label{sec:exponential}

\begin{figure}[ht]
\centering
\includegraphics[width=0.45\textwidth]{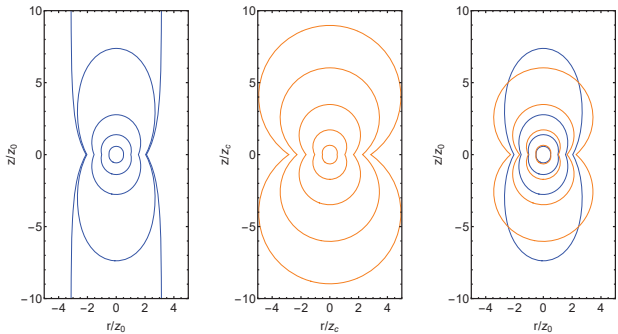}
\caption{
Shock fronts of bubble for different times $y$. {\it Left panel}: Evolution of a bubble in a halo of exponential density profile.
Different contours denote ${\tilde y}=y/z_0=0.5,\,1.0,\,1.5,\,1.95,\,2.0$.
{\it Middle panel}: Evolution of a bubble in in a halo of power-law density profile.
Different contours denote ${\tilde y}=y/z_c=0.5,\,1.0,\,1.5,\,1.95,\,2.3$.
{\it Right panel}: Comparison of bubbles in exponential halo (in blue) and power-law halo (in orange) for different times
(${\tilde y}=0.5,\,1.0,\,1.5,\,1.95$).
The two scale heights are set to be the same ($z_0=z_c$).
The bubble in the power-law halo is rounder.
The top of the blue bubble catches up with the orange bubble and extends to infinity in finite time.
}
\label{fig:evolve}
\end{figure}

For an exponential density profile of the halo, ${\cal R}(z)=\exp\left(-z/z_0\right)$, the solution to Equation~(\ref{eq:kompaneets})
is the classic Kompaneets solution,
\begin{equation}\label{eq:exponential}
  \cos\left(\frac{r}{2z_0}\right)=\frac{1}{2\sqrt{{\cal R}}}\left({\cal R}+1-\frac{y^2}{4z_0^2}\right)\,.
\end{equation}
The top of the bubble is given by
\begin{equation}\label{eq:top_exponential}
  \sqrt{{\cal R}_u}=\exp\left(-\frac{z_u}{2z_0}\right)=1-\frac{y}{2z_0}\,,
\end{equation}
and its velocity
\begin{equation}\label{eq:vel_top_exponential}
  v_u=\frac{d z_u}{d t}=\exp\left(\frac{z_u}{2z_0}\right)\, \frac{d y}{d t}=\frac{1}{\sqrt{{\cal R}_u}}\, \frac{d y}{d t}\,.
\end{equation}
The left panel of Figure~\ref{fig:evolve} shows the development of the bubble in an exponential halo.
Different contours denote the boundary (i.e., the shock) of the bubble at different $y$ (i.e., different times).
In an exponential halo, the bubble will developed into a structure asymptotically similar to a cylinder in finite time.
As $y\to 2z_0$, the top of the bubble $z_u\to\infty$, and the lateral radius of the bubble $r(z,y\to 2z_0)\to 2z_0\cos^{-1}(\sqrt{{\cal R}}/2)$
(e.g., at the base of the halo $r\to 2\pi z_0/3$, and high up in the halo $r\to \pi z_0$.)

If the shock velocity at any $z$ drops below the sound speed $c_{\rm s}$, then it decays and is absorbed in the halo gas.
Otherwise, a shock is able to penetrate into the halo with a velocity higher than the sound speed $c_{\rm s}$
and transfers the energy of the initial central source into the exponential atmosphere.
\citet{baumbr13} defined the condition of shock penetration into the exponential atmosphere
when the velocity of shock front is higher than $c_{\rm s}$,
$v_u>3 c_{\rm s}$.

\begin{figure*}[]
\centering
\plotone{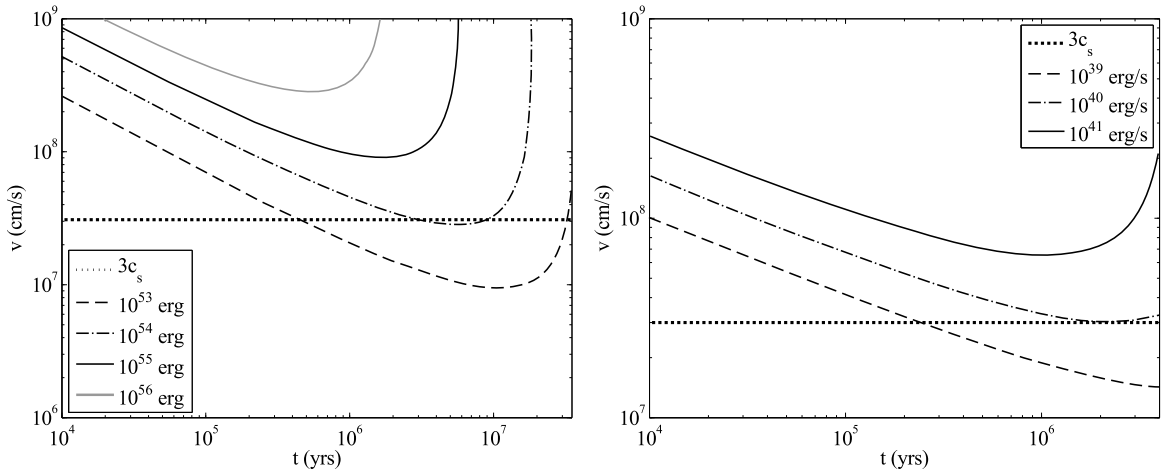}
\caption{
Temporal variation of the shock velocity of the top of the bubble for the case of exponential halo with $z_0=0.67$ kpc and $n_0=0.03$ cm$^{-3}$.
{\it Left panel}: One single input of energy at the GC.
{\it Right panel}: Multiple TDEs with different values of the power release in the GC.
The horizontal dotted line indicates the velocity which is necessary for the shock in order not to stall in the halo.
It is three times the sound speed in the halo $3\times 10^7$ cm s$^{-1}$.
}
\label{exponential}
\end{figure*}

For illustration purpose we show in Figure~\ref{exponential} (the left panel)
the velocity distribution for the case $z_0=0.67$ kpc and $n_0=0.03$ cm$^{-3}$,
with single input of energy $W=10^{56}$ erg (thin solid line), $W=10^{55}$ erg (thick solid line),
$W=10^{54}$ erg (dash-dotted line), and $W= 10^{53}$ erg (dashed line).
The dotted line shows the level $3c_{\rm s}$ in the halo $3\times 10^7$ cm s$^{-1}$.


For a single source the energy occupies more and more volume of the exponential atmosphere at $t>0$,
eventually reaching a top at infinity at finite time.
This is an artefact of the Kompaneets' solution, as the shock speed cannot exceed the speed of light.
In the end, the structure of a single source disappears as the energy it releases is distributed over an infinite volume.

For the parameters in the GC a single star disruption is unable to provide enough energy for
the FBs or similar structures,
i.e., no more than $10^{52}\sim 10^{53}$ erg \citep{dai18}.
Our calculations show that an unbelievably single star event with an energy
exceeding $W=10^{54}$ erg
is needed to penetrate the disk into the halo.

Alternatively this energy can be supplied by a series of many weaker disruption events with an effective luminosity ${\dot{W}}$.
It follows from \citet{dai18} that any event of star disruption provides an energy of $10^{52}\sim 10^{53}$ erg
with the average rate of star capture about $10^{-4}$ yr$^{-1}$.
We show the temporal velocity variations for different values of ${\dot{W}}$ in the initial development of the structure in the halo
in the right panel of Figure~\ref{exponential}.
We show that the velocity of the envelope exceeds the sound speed if ${\dot{W}}\ga 10^{40}$ erg s$^{-1}$.
As a result
an outflow in the halo can be provided by normal successive star disruptions.

\subsection{Power-law Density Profile}
\label{sec:powerlaw}

We mimic the distribution described by Equation~(\ref{eq:beta-model}) (or (\ref{beta_atm})) by a power-law density profile,
${\cal R}(z)=\left(1+z/z_c\right)^{-3\beta}$. These profiles are similar, in particular, for $z\gg z_c$.
If $\beta=2/3$, the solution to Equation~(\ref{eq:kompaneets}) for the power-law density profile is,
\begin{equation}\label{eq:powerlaw}
  \left(\frac{r}{z_c}\right)^2=\sinh^2\left(\frac{y}{z_c}\right) - \left[1+\frac{z}{z_c} - \cosh\left(\frac{y}{z_c}\right)\right]^2 \,.
\end{equation}
The top of the bubble is given by
\begin{equation}\label{eq:top_powerlaw}
  \frac{z_u}{z_c}=\cosh\left(\frac{y}{z_c}\right)+\sinh\left(\frac{y}{z_c}\right)-1\,,
\end{equation}
and its velocity
\begin{equation}\label{eq:vel_top_powerlaw}
  v_u=\frac{d z_u}{d t}=\left(1+\frac{z_u}{z_c}\right)\, \frac{d y}{d t}=\frac{1}{\sqrt{{\cal R}_u}}\, \frac{d y}{d t}\,.
\end{equation}
The middle panel of Figure~\ref{fig:evolve} shows the development of the bubble in a power-law halo.
In such a halo, the bubble behaves like an ascending and expanding sphere rising from the base.
In contrast to the bubble in an exponential halo, the top of the bubble (or any other part) cannot reach infinity in finite time.

\begin{figure*}[]
\centering
\plotone{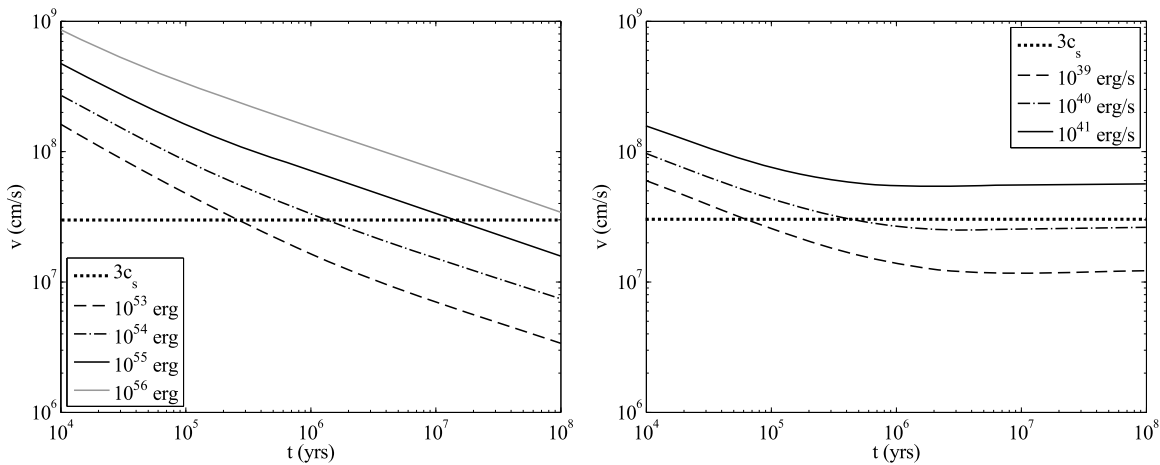}
\caption{
Same as Figure~\ref{exponential} except for power-law halo with $z_c=0.26$ kpc and $n_0=0.46$ cm$^{-3}$.
{\it Left panel}: One single input of energy at the GC.
{\it Right panel}: Multiple TDEs with different values of the power release in the GC.
The horizontal dotted line is three times the sound speed in the halo $3\times 10^7$ cm s$^{-1}$.
}
\label{beta}
\end{figure*}

For the single explosion the lifetime of the envelope is restricted in the halo
(even for energy as high as $W= 10^{56}$ erg)
and its velocity decreases continuously,
see left panel of Figure~\ref{beta}.


For the case of continuous multi-captures for different values of power, the outflow velocity is shown in the right panel of Figure~\ref{beta}.
We have shown that the velocity of the envelope is constant in the atmosphere and exceeds the sound speed if ${\dot{W}}\ga 10^{41}$ erg s$^{-1}$.

When we compare the bubbles in these two density profile, the shock envelope in a power-law halo becomes broader than in the exponential halo.
Therefore the top of the bubble will have travelled a smaller distance from the plane for similar bubble volumes.
Apart from that the expansion is quite comparable, and since we expect such a density decrease to hold only for the lower halo,
subsequently crossing over to an exponentially decreasing halo at larger distances,
the main difference will be a somewhat larger energy input at the base of the halo in order to compensate for this effect.

%
%
%
%
%

\section{Numerical results}
\label{sec:simulation}

To consolidate the findings of Section~\ref{sec:analytical} we perform hydrodynamic simulations of single and sequential multiple
``explosions'' in an atmosphere or a halo.
An ``explosion'' means a burst of energy input representing the energy release of a TDE.
We adopted the publicly available MHD simulation package FLASH \citep{fryx00,fryx10}.
To reflect the simple situation in Section~\ref{sec:analytical},
we keep only the minimal physics (and do not consider magnetic field, heating and cooling, etc.).

The background halo (or the initial condition for the halo) is an isothermal gas (i.e., pressure is proportional to density)
in hydrostatic equilibrium (i.e., gravity is balanced by pressure gradient).
We consider two types of density distributions: (1) exponential, see Equation~(\ref{eq:exp-model}),
and (2) $\beta$-model, see Equation~(\ref{eq:beta-model}) (with $R=0$).
Once the density profile is known, the gravitational field required for hydrostatic equilibrium can be obtained.
In fact, the sole purpose of the gravitational field here is to keep the unperturbed background in equilibrium,
it does not affect the development of the bubbles as the thermal and kinetic energy of the bubble is much larger than the potential energy.

An explosion is implemented as an instantaneous energy release in the form of
pure thermal energy uniformly distributed in a sphere of radius $\sim 200$ pc.

\begin{figure}[ht]
\centering
\plotone{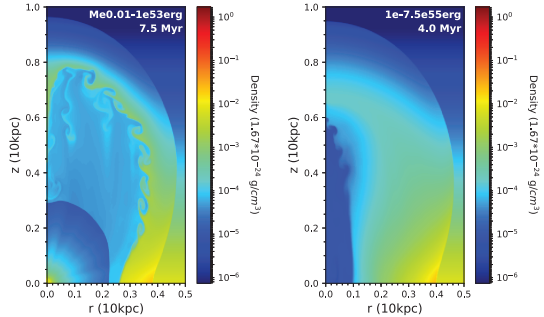}
\caption{
Numerical simulation of the Fermi Bubbles
or similar structures
in an exponential halo.
The color maps in the figure are the density distribution in logarithmic scale.
{\it Left panel}: Multiple TDEs with each TDE releasing $10^{53}$ erg of energy and the interval between successive TDEs is $0.01$ Myr;
the simulation ends at 7.5 Myr.
{\it Right panel}: Single TDE with an energy release $7.5\times 10^{55}$ erg; the simulation ends at 4 Myr.
The total energy released at the end of the simulations is the same for the two cases, i.e., $7.5\times 10^{55}$ erg.
}
\label{log_density_exp_atm_samescale}
\end{figure}

Figure~\ref{log_density_exp_atm_samescale} shows the density distribution of the bubble in an exponential halo.
The temperature of the halo and the mass density at the mid-plane are taken as $10^6$ K and $1.24\times 10^{-26}$ g cm$^{-3}$.
The left panel shows the case of sequential multiple TDEs or explosions
that each explosion releases $10^{53}$ erg and the interval between explosions is 0.01 Myr.
This provides an average power (or luminosity) ${\dot{W}}=3.3\times 10^{41}$ erg s$^{-1}$.
The simulation stops at 7.5 Myr as the bubble boundary approaches the edge of the simulation box.
For comparison, in the right panel, we show the case of a single TDE or explosion.
We take the energy release of this explosion to be $7.5\times 10^{55}$ erg,
which is the same as the total energy release of the case of multiple explosions at the end of its simulation.
However, in the single explosion case the expansion rate is higher, and the bubble is of comparable size
(the top is $9\sim 10$ kpc) at 4 Myr only.

\begin{figure}[ht]
\centering
\plotone{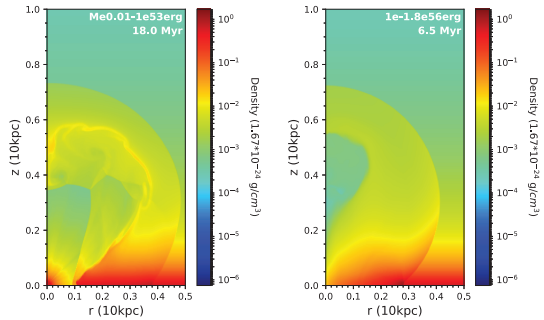}
\caption{
Same as Figure~\ref{log_density_exp_atm_samescale} except for a $\beta$-model halo.
The color scale is the same as in Figure~\ref{log_density_exp_atm_samescale}.
{\it Left panel}: Multiple TDEs with each TDE releasing $10^{53}$ erg of energy and the interval between successive TDEs is $0.01$ Myr;
the simulation ends at 18 Myr.
{\it Right panel}: Single TDE with an energy release $1.8\times 10^{56}$ erg; the simulation ends at
6.5 Myr.
The total energy released at the end of the simulations is the same for the two cases, i.e., $1.8\times 10^{56}$ erg.
}
\label{log_density_beta_atm_samescale}
\end{figure}

Figure~\ref{log_density_beta_atm_samescale} shows the density distribution of the bubble in a $\beta$-model halo.
The temperature of the halo and the density at the mid-plane are taken as $1.26\times 10^6$ K and
$9.50\times 10^{-25}$ g cm$^{-3}$,
and the parameter $\beta=0.71$ \citep{mill13}.
The left panel shows the case of sequential multiple explosions
that each explosion releases $10^{53}$ erg and the interval between explosions is 0.01 Myr.
The simulation stops at 18 Myr.
In the right panel, we show the case of a single explosion with the energy release $1.8\times 10^{56}$ erg,
same as the total energy release of the case of multiple explosions at the end of its simulation.
In the single explosion case the expansion rate is higher, we stop the simulation at
6.5 Myr
when the bubble is of comparable size.

Comparing the two examples presented here,
we note that it takes more time and energy for the bubble to develop in the $\beta$-model halo than the exponential halo.

From Figures~\ref{log_density_exp_atm_samescale} \& \ref{log_density_beta_atm_samescale},
we observed that the bubble envelope is more slender in an exponential halo while it is rounder in a $\beta$-model halo.
(Note that the color scales of Figures~\ref{log_density_exp_atm_samescale} \& \ref{log_density_beta_atm_samescale} are the same.)
These agree well with the analytical results in Section~\ref{sec:analytical}.
Moreover, the interior of the bubble in the case of multiple explosions has a lot of shocks and is more turbulent for both halos.
Thus, the routine explosions or routine TDEs scenario is more conducive to particle acceleration.

\section{Discussion and Conclusions}
\label{sec:summary}

%
We have shown by analysing the energy release from TDEs,
it is possible to provide a sufficient amount of energy, even for a molecular cloud environment near the GC.
The crucial point here is that the filling factor of molecular gas is sufficiently  below unity, as is the case for the normal ISM.
We have shown that density profiles following an exponential decay or a $\beta$-model in the lower halo cannot stall shocks in general.
In addition, as we have shown, the ongoing energy input by TDEs resembles more that of a wind than a point explosion,
which can drive the shock further out, as energy is constantly added to the bubble thus delaying catastrophic cooling by reheating,
as has been shown by \citet{fk98} in the context of the Galactic fountain.


Here is a summary of our conclusions.
\begin{itemize}
\item The standard interpretation of the FB origin is one huge energy release about $10^{55} \sim 10^{56}$ ergs in the GC \citep[see e.g.,][]{su10,zub12},
whose envelope propagates through the Galactic halo.
Our calculations show that, for the case of $\beta$-model halo density profile,
the envelope disappears in the halo within $10^8$ yrs even for an energy as high as $10^{56}$ erg (see left panel of Figure~\ref{beta}).
For the case of exponential density profile, the energy release must exceed $10^{54}$ erg in order for the envelope to penetrate into the halo
(see left panel of Figure~\ref{exponential}).

\item \citet{cheng11} suggested a phenomenological model of FBs as a result of routine star disruptions by the supermassive black hole in the GC.
They did not derive the model parameters quantitatively, but had a rough estimation of the energy release $\ga 10^{53}$ ergs.
In our present analysis we concluded that the expected energy release is indeed about $10^{53}$ ergs.
\citet{dai18} showed that the accretion energy is mainly carried away by radiation ($\sim 3$\%), a jet ($\sim 20$\%), hydrodynamic outflow ($\sim 20$\%).
The total energy of the outflow is about $10^{52} \sim 10^{53}$ ergs, which is about 10\% of the total rest mass energy of the disrupted star.
In addition, if the X-ray chimney near GC \citep[of energy $\sim 4\times 10^{52}$ ergs,][]{ponti19}
and radio bipolar bubbles near the GC \citep[of energy $\sim 7\times 10^{52}$ ergs,][]{hey19} come from a TDE by the supermassive black hole at the GC,
then it is consistent with our estimates concerning the energy release by TDEs.

\item The main concern is that the energy release for one TDE of $10^{53}$ ergs is not high enough to penetrate through a medium of 50 cm$^{-3}$ around the GC.
However, routine star captures
at a rate $\ga 10^{-4}$ yr$^{-1}$
resulting in an average energy luminosity at the GC of about
$\dot{W} \sim 3\times 10^{41}$ erg s$^{-1}$
can enable a shock penetrate through the disk into the halo
and form the FBs or similar structures.

\item Another concern is that the gas of multi-component distribution in the halo is not well-known.
We considered two models of gas distribution in the halo: exponential and beta-model.
The evolution of FB envelope
(and similar structures)
is quite different for the two models.

For the exponential distribution the envelope propagates through the halo with acceleration.
A stationary envelope structure is formed at a finite time (of the order of $10^7$ yrs),
and the lateral radius of the envelope is about $\pi$ times the scale height.
Moreover, the top of the envelope may be destroyed by Rayleigh-Taylor instability \citep[see][]{baumbr13}.

For the $\beta$-model halo, the envelope propagates through the halo with deceleration.
In the limit, the velocity at the top is constant, and the envelope can extend to infinity in all directions.

\item Thus, we conclude that for routine TDEs at the GC, the bubble can exist for a long time ($>10^7$ yrs) provided that
$\dot{W} > 10^{40}\sim 10^{41}$ erg s$^{-1}$ (see right panels of Figures~\ref{exponential} \& \ref{beta}).

\item Numerical simulations agree with the analytical solutions for both exponential and $\beta$-model halos.
From the two examples we have in Section~\ref{sec:simulation}, it takes more energy and time for the bubble to develop in a $\beta$-model halo.
Moreover, when comparing with the single huge energy release case, the interior of the bubble in the multiple TDEs' case
is more turbulent and is more promising for cosmic ray acceleration.

\item The presented hydrodynamic analysis may serve as background models for acceleration processes in the FBs
or similar structures.
\end{itemize}

\section*{Acknowledgments}

The authors are grateful to Katia Ferri\`{e}re, Miguel Avillez and Takeshi Oka for fruitful discussions.
CMK and HC are supported in part by the Taiwan Ministry of Science and Technology grants MOST 105-2112-M-008-011-MY3,
MOST 108-2112-M-008-006 and MOST 109-2112-M-008-005.
DOC and VAD are supported by the grants RFBR 18-02-00075 and RSCF 20-12-00047.
DOC is supported in parts by foundation for the advancement of theoretical physics ``BASIS''.
LD acknowledges support by the Danish National Research Foundation (DNRF132).

\software{FLASH
\citep[http://flash.uchicago.edu/site/flashcode/][]{fryx00,fryx10}.
}

\end{document}